\documentclass[aps, reprint,prl,superscriptaddress,nofootinbib,floatfix]{revtex4-2}
\usepackage{dcolumn}
\usepackage[utf8]{inputenc}
\usepackage{amsmath}
\usepackage{amsfonts}
\usepackage{amssymb}
\usepackage{mathtools}
\usepackage{graphicx}
\usepackage{hyperref}
\usepackage{xcolor}
\usepackage{comment}
\usepackage[toc,page]{appendix}
\usepackage[english]{babel}
\usepackage{tikz}
\usetikzlibrary{quantikz2}

\newcommand{\CNOT}{\textsc{cnot}}

\newcommand{\rmi}{\mathrm{i}}

\newtheorem{teorema}{Theorem}

\newcommand{\nuc}[2]{\ensuremath{^{#1}\textrm{#2}}}

\begin{document}
\title{Temporal Entanglement and Witnesses of Non-Classicality}

\author{Giuseppe Di Pietra}
\email{giuseppe.dipietra@physics.ox.ac.uk}
\affiliation{Clarendon Laboratory, University of Oxford, Parks Road, Oxford OX1 3PU, United Kingdom}

\author{Gaurav Bhole}
\affiliation{Department of Cancer Biology, Dana-Farber Cancer Institute, Boston, MA, USA 02215}
\affiliation{Department of Biological Chemistry and Molecular Pharmacology, Harvard Medical School, Boston, MA, USA 02115}
\affiliation{Department of Physics, Faculty of Arts and Sciences, Harvard University, Cambridge, MA, USA 02138}

\author{James Eaton}
\affiliation{Physical and Theoretical Chemistry, University of Oxford, South Parks Road, Oxford OX1 3QZ, United Kingdom}
\affiliation{Kavli Institute for Nanoscience Discovery, University of Oxford, Sherrington Road, Oxford OX1 3QU, United Kingdom}

\author{Andrew J. Baldwin}
\affiliation{Physical and Theoretical Chemistry, University of Oxford, South Parks Road, Oxford OX1 3QZ, United Kingdom}
\affiliation{Kavli Institute for Nanoscience Discovery, University of Oxford, Sherrington Road, Oxford OX1 3QU, United Kingdom}

\author{Jonathan A. Jones}
\affiliation{Clarendon Laboratory, University of Oxford, Parks Road, Oxford OX1 3PU, United Kingdom}

\author{Vlatko Vedral}
\affiliation{Clarendon Laboratory, University of Oxford, Parks Road, Oxford OX1 3PU, United Kingdom}

\author{Chiara Marletto}
\affiliation{Clarendon Laboratory, University of Oxford, Parks Road, Oxford OX1 3PU, United Kingdom}

\date{\today}%

\begin{abstract}
The universality of quantum theory has been questioned ever since it was proposed. Key to this long-unsolved question is to test whether a given physical system has non-classical features. Here we connect recently proposed witnesses of non-classicality, based on information-theoretic ideas, with the theory of temporal entanglement. We provide a protocol to witness the non-classicality of a system by probing it with a qubit: we show that, assuming a general conservation law, violating temporal Bell inequalities on the qubit probe implies the non-classicality of the system under investigation. We also perform proof-of-principle experimental emulations of the proposed witness of non-classicality, using a three qubit Nuclear Magnetic Resonance quantum computer. Our result is robust, as it relies on minimal assumptions, and remarkably it can be applied in a broad range of contexts, from quantum biology to quantum gravity.
\end{abstract}

\maketitle
\section{Introduction}
Quantum theory stands as one of the best available explanations in contemporary physics, providing a robust description of physical phenomena within its domain of applicability and introducing profound conceptual innovations such as superposition and entanglement -- phenomena with no classical analogues.
However, a fundamental debate persists regarding the \textit{universality} of quantum theory, specifically the question of whether its formalism can be extended to encompass all physical systems, at any scale. This dispute traces its origins to Heisenberg, Bohr and Von Neumann's studies  \cite{heisenberg_physical_1949,bohr_quantum_1928,von_neumann_mathematische_1996}, and is further amplified when attempting to reconcile quantum theory with the other most successful theory of modern physics, the completely classical framework of general relativity \cite{marletto_why_2017}, or when considering the question of whether quantum effects could play a role even in macroscopic bio-molecules, for instance to explain the remarkable efficiency of some biological processes \cite{kim_quantum_2021}.

{All such diverse issues have one element in common. They consider the possibility of a \textit{hybrid system}, defined here as a composite system comprising a quantum system $Q$ (e.g. a quantum mass or a photon) interacting with a ``mysterious" system $M$ (e.g. gravity, a macroscopic biosystem, or measuring apparatus), whose dynamics may deviate from those of quantum theory, for example by being fully classical.
Several approaches have been developed to investigate hybrid systems: from arguments challenging their consistency \cite{page_indirect_1981,deutsch_quantum_1985,terno_inconsistency_2006,marletto_quantum_2022}, to models describing the emergence of classicality \cite{diosi_universal_1987,diosi_models_1989,penrose_gravitys_1996,zurek_decoherence_2003,kafri_classical_2014}, to hybrid quantum-classical formalisms \cite{sherry_interaction_1978,diosi_quantum_2000,peres_hybrid_2001, hall_interacting_2005}.}

More recently, a new direction has emerged in the field, aimed at designing tests to check whether $M$ has quantum properties, by using the quantum system $Q$ as a probe. Here by ``quantum properties" we shall mean, more precisely, ``non-classicality", which we define as the existence of (at least) two non-commuting variables characterising $M$. We call these tests \textit{witnesses of non-classicality} (see \cite{marletto_witnessing_2020}). If $M$ were strictly classical -- that is, if its physical variables were all commuting -- then $M$ would not be capable of mediating certain physical transformations on $Q$, called \textit{witnessing tasks}. Thus, if $M$ can mediate the witnessing tasks on $Q$, one can conclusively infer the quantum-mechanical nature of $M$ \cite{marletto_quantum-information_2025}. Crucially, one would like to probe $M$ in a way that is (i) independent of the particular dynamics governing the hybrid system, and (ii) doesn't require direct experimental access to or control of $M$ itself, but only to let $M$ interact with $Q$ after initialisation in a suitable state.

An example of such witnesses is the recently proposed \textit{temporal witness of non-classicality}, which exploits the \textit{temporal} evolution of the quantum probe $Q$ as a witnessing task to infer the quantum properties of its less-accessible partner $M$ \cite{di_pietra_temporal_2023}. Specifically:
\begin{teorema}
    \textbf{Temporal Witness of Non-Classicality} \\
    If $M$ can induce a quantum coherent evolution of $Q$, then $M$ must be non-classical, provided that a global observable of the system $Q\oplus M$ is conserved.
    \label{theorem:temporalold}
\end{teorema} In other words, if the probe $Q$ exhibits dynamics that defy classical explanation, this serves as indirect evidence that $M$ possesses intrinsically quantum properties as defined above.
The key assumption of Theorem~\ref{theorem:temporalold} is the \textit{conservation} of a global observable of the composite system $Q\oplus M$, for example, its \textit{total energy} or \textit{momentum}. 
An application of this theorem to biological systems was proposed in \cite{di_pietra_temporal_2024}.

Another example of a witness of non-classicality, known as the \textit{General Witness Theorem} (GWT), exploits the \textit{spatial} entangling capacity of the unknown system $M$ as a witnessing task to its quantum features \cite{marletto_witnessing_2020}:
\begin{teorema}
    \textbf{General Witness Theorem} \\
    If $M$ can entangle $Q$ and $Q'$, by locally interacting with each of them, then $M$ is non-classical
    \label{theorem:gwt}
\end{teorema} which was notably applied to gravity in \cite{bose_spin_2017,marletto_gravitationally_2017}. 
The GWT offers a very robust argument to infer $M$'s non-classicality, as it relies on two general assumptions (i) the Principle of Locality \cite{di_pietra_role_2024} and (ii) the Principle of Interoperability of Information \cite{deutsch_constructor_2015}, in a completely dynamics- and scale-agnostic setting. 
Despite the theoretical robustness of the GWT, its experimental implementation presents significant challenges, because it requires two quantum probes that must be \textit{space-like separated} for the generation of spatial entanglement to constitute a meaningful witnessing task \cite{di_pietra_bose-marletto-vedral_2024}.
In contrast, the temporal witness (Theorem~\ref{theorem:temporalold}) works with a single probe  and thus appears simpler to realise, even if it is currently formulated in a less general framework than the GWT in Theorem~\ref{theorem:gwt}. 

The spatial witness~\ref{theorem:gwt} has the advantage of using ``$M$-mediated" entanglement generation on the two quantum probes to reveal $M$'s non-classicality. In this paper we show how the temporal witness~\ref{theorem:temporalold} can be expressed as using the ``$M$-mediated" {\sl temporal entanglement} on the probe $Q$ to infer $M$'s non-classicality. We shall demonstrate this novel \textit{temporal-entanglement-based} witness of non-classicality using the same assumptions as for Theorem~\ref{theorem:temporalold}; then we shall assess if the temporal Bell inequality \cite{brukner_quantum_2004} on the single quantum probe $Q$ can be violated when its dynamical evolution is mediated by (1) a classical $M$ and (2) a quantum $M$. 
We show that \textit{only a non-classical system} is capable of inducing a violation of the temporal Bell inequality for $Q$, and thus to \textit{entangle in time} $Q$ at times $t_0$ and $t>t_0$. This establishes a surprising connection between witnesses of non-classicality and the theory of temporal entanglement.

We also perform an experimental emulation of this novel witness using three Nuclear Magnetic Resonance (NMR) qubits. The unknown system $M$ in this emulation would be the first qubit, interacting with the quantum probe $Q$ represented by the second qubit. The non-classicality of $M$ will be simulated by controlling the possible interactions with $Q$. The third qubit $A$ is an ancilla introduced to record the results of measurements done on $Q$ at times $t_0$ and $t>t_0$. The experiment confirms that mediated \textit{temporal} entanglement disappears when considering (1) a classical $M$ and (2) a quantum $M$ prepared in an eigenstate of the classical observable. Our results bridge the conceptual gap between the GWT and the temporal witness of non-classicality, establishing the \textit{capacity to induce entanglement on $Q$} -- both spatial and temporal -- as the definitive witnessing task for inferring $M$'s non-classicality.
Furthermore, our analysis connects the assumptions of the two Theorems~\ref{theorem:temporalold} and ~\ref{theorem:gwt}, suggesting a formulation of the so-called \textit{locality in time} in terms of \textit{energy-conservation}. 
Finally, this investigation can provide a rigorous framework to extend Leggett--Garg tests beyond the single system scenario \cite{emary_leggettgarg_2013}. 

\section{Temporal Entanglement-based witness of non-classicality}
The temporal Bell inequality, derived from the assumptions of \textit{c}-number realism and locality \textit{in time}, determines a bound on the temporal correlations $E(A_i,B_j)$ between successive measurements of a pair of observables $A_i(t_0)$, $B_j(t)$, $i,j=1,2$ on a physical system $Q$, when these correlations can be described by a classical joint probability distribution \cite{brukner_quantum_2004, fritz_quantum_2010}:
\begin{equation}
    \mathcal{B}\equiv\left|E(A_1,B_1)-E(A_1,B_2)+E(A_2,B_1)+E(A_2,B_2)\right|,
    \label{eq:temporalBellineq}
\end{equation} the bound being $\mathcal{B}\leq 2$. A violation of the inequality rejects a \textit{c}-number, local \textit{in time} description of reality in favour of a \textit{q}-number based one, as for the spatial Bell inequality \cite{di_pietra_role_2024},  introducing the notion of \textit{entanglement in time} \cite{brukner_quantum_2004}.

\begin{figure}
    \centering
    \includegraphics[width=\linewidth]{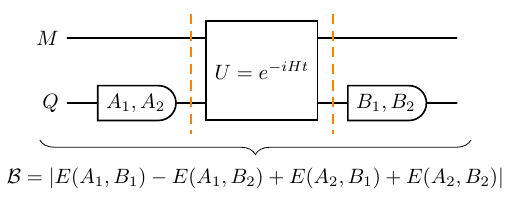}
    \caption{Schematic description of the process to compute the temporal Bell quantity $\mathcal{B}$. Two chosen variables of the quantum probe $Q$, $A_1$ and $A_2$, are measured at time $t_0$. The hybrid system $Q\oplus M$ evolves under the global energy-conserving dynamics generated by $H$. The two time-evolved variables $B_1=U^\dagger A_1 U$ and $B_2=U^\dagger A_2 U$ are then measured at time $t>t_0$. Repeating this experiment allows us to construct $\mathcal{B}$ and check whether it violates the classical boundary $\mathcal{B}=2$. This will be a witness to $M$'s non-classicality.}
    \label{fig:CHSHscheme}
\end{figure}

When considering a hybrid system $Q\oplus M$, with projective measurements performed on $Q$ only, the temporal correlation function reads:
\begin{equation}
    E(A,B)=\frac{1}{2}\text{Tr}\left(AB\rho_M\right)
    \label{eq:corrfunctionhybrid}
\end{equation} and it depends on the chosen observables measured on the quantum probe $Q$, but crucially also on the \textit{mediator's initial state} $\rho_M$: whether the quantum probe $Q$ violates the temporal Bell's inequality depends on the initial state of the unknown system $M$ inducing its temporal evolution. 
We stress here that no measurements are performed on the unknown system $M$, as in general we cannot access it: this is the system we want to investigate thanks to the witness of non-classicality, looking at its capability of \textit{mediating} temporal quantum correlations on $Q$ under an energy conserving dynamics. We shall call the unknown system $M$ the \textit{``mediator"} in what follows.
Moreover, Eq.~\ref{eq:corrfunctionhybrid} does not depend on the initial state of the quantum probe $Q$, which allows us to adopt the Heisenberg picture to include the temporal evolution of the hybrid system between the two measures. Specifically, the evolution must be \textit{energy-conserving}, to comply with the assumptions of the temporal witness: $B=U^\dagger A U=e^{\rmi Ht} A e^{-\rmi Ht}$ where $[H,E_{tot}]=0$, with $E_{tot}$ being the total energy of the hybrid system, see Fig.~\ref{fig:CHSHscheme}. For example, considering a hybrid system made up by a qubit $Q$ and a bit $M$, the total energy will be $E_{tot}=a\sigma_z \otimes \mathrm{I}+b\mathrm{I}\otimes \sigma_z $, where $\sigma_z$ is the Pauli operator, $\mathrm{I}$ is the unit operator and $a,b\in\mathbb{R}$.

We shall now show that a violation of this inequality, when considering an energy-conserving dynamics for $Q\oplus M$, is a witness of $M$'s non-classicality.

Without loss of generality, we shall consider the quantum probe $Q$ to be a qubit whose algebra is generated by the descriptors $\hat{\mathbf{q}}^Q=\left\{q_x^Q,q_y^Q,q_z^Q\right\}=\left\{\sigma_x\otimes \mathrm{I},\sigma_y\otimes \mathrm{I},\sigma_z\otimes \mathrm{I}\right\}$. $M$ is instead a \textit{classical} bit, whose algebra is  generated by the descriptor $\hat{q}^M=q_z^M=\mathrm{I}\otimes \sigma_z$. 
The state of the quantum probe $Q$ at time $t$ is completely characterised by at least two components, for example, $\{q_x^Q,q_z^Q\}$. For this reason, we choose to measure the observables $A_1=q_z^Q$, $A_2=q_x^Q$ at time $t_0$. The hybrid system constructed in this way evolves according to the most general energy-conserving Hamiltonian:
\begin{equation}
    H_{CM}=aq_z^Q+bq_z^M+cq_z^Qq_z^M+r\mathrm{I_2},
    \label{eq:HamiltonianClassical}
\end{equation} where $a,b,c\in\mathbb{R}$ and $\mathrm{I_2}=\mathrm{I}\otimes\mathrm{I}$, and we measure the observables $B_1=e^{\rmi H_{CM}t}q_z^Qe^{-\rmi H_{CM}t}$ and $B_2=e^{\rmi H_{CM}t}q_x^Qe^{-\rmi H_{CM}t}$ to compute the Bell quantity in Eq.~\ref{eq:corrfunctionhybrid}:
\begin{equation}
    \mathcal{B}=\frac{1}{2}\left|2+\cos{\left[2(a-c)\right]}+\cos{\left[2(a+c)\right]}\right|\leq2.
    \label{eq:Bclassicalmediator}
\end{equation}
The temporal Bell quantity $\mathcal{B}$ computed for the quantum probe $Q$ when its evolution is induced by a \textit{classical mediator} $M$ never exceeds the classical boundary $\mathcal{B}=2$. Said differently, a classical system $M$ cannot create temporal quantum correlation in the evolution of a quantum system, meaning that its evolution must be classical, i.e., describable with a joint classical probability distribution. Thus, observing a quantum global energy-conserving evolution of the quantum probe $Q$ induced by an unknown mediator $M$ is a witness to its non-classicality, proving that $M$ is \textit{entangling in time} the quantum probe $Q$ at times $t_0$ and $t>t_0$.

Let us show now that this is the case by considering a \textit{quantum} mediator $M$. Without loss of generality, $M$ is a qubit with descriptors $\hat{\mathbf{q}}^M=\left\{q_x^M,q_y^M,q_z^M\right\}=\left\{\mathrm{I}\otimes\sigma_x,\mathrm{I}\otimes\sigma_y,\mathrm{I}\otimes\sigma_z\right\}$. The most general global energy-conserving Hamiltonian will thus read:
\begin{align}
    H_{QM}=&aq_z^Q+bq_z^M+cq_z^Qq_z^M+f\left(q_x^Qq_x^M+q_y^Qq_y^M\right) & \nonumber \\ & +g\left(q_x^Qq_y^M-q_y^Qq_x^M\right)+r\mathrm{I}_2;
    \label{eq:HamiltonianNonClassical}
\end{align} we choose to measure again $A_1=q_z^Q, A_2=q_x^Q$ at time $t_0$, and $B_1=e^{\rmi H_{QM}t}q_z^Qe^{-\rmi H_{QM}t},B_2=e^{\rmi H_{QM}t}q_x^Qe^{-\rmi H_{QM}t}$ at time $t>t_0$.
Recalling the definition of the temporal correlation function for hybrid systems given in Eq.~\ref{eq:corrfunctionhybrid}, we must distinguish two cases depending on the initial state of the mediator $M$. 
Since the mediator $M$ cannot be accessed -- it is the unknown sector of the hybrid system -- the initial state $\rho_M$ will be determined by previous interactions it had with the environment, which are not relevant to the protocol we are interested in.

\textbf{Classical initial state of $\mathbf{M}$}.-- We shall define a \textit{classical} state as one that commutes with the global conserved quantity: $\left[\rho_M^{class}, E_{tot}\right]=0$. Thus, classical initial states for $M$ are the eigenstates of its $Z$ component, such as $\rho_M^{class}\equiv \rho_M^{\pm}=\frac{1}{2}\left(\mathrm{I}\pm q_z^M\right)$, or a mixed state, for example, the maximally mixed one $\rho_M^{class}\equiv\rho_M^{mix}=\mathrm{I}/2$.
In the former case, $\mathcal{B}$ is:
\begin{align}
    &\mathcal{B}^{\pm}=\left|\cos{\left(\left(a+b\pm2c\right)t\right)}\cos{\left(\sqrt{\left(a-b\right)^2+4\left(f^2+g^2\right)}t\right)} \right. \nonumber & \\ & \left. +\frac{\left(a-b\right)^2+4\left(f^2+g^2\right)\cos^2{\left(\sqrt{\left(a-b\right)^2+4\left(f^2+g^2\right)}t\right)}}{\left(a-b\right)^2+4\left(f^2+g^2\right)} \right. \nonumber & \\ & \left. - \frac{\left(a-b\right)\sin{\left(\left(a+b\pm2c\right)t\right)}\sin{\left(\sqrt{\left(a-b\right)^2+4\left(f^2+g^2\right)}t\right)}}{\sqrt{\left(a-b\right)^2+4\left(f^2+g^2\right)}}\right|
    \label{eq:Bellclassicalstate}
\end{align}
while in the latter case $\mathcal{B}^{mixed}=(\mathcal{B}^++\mathcal{B}^-)/2$, due to the linearity of the trace.
In both cases, the Bell quantity $\mathcal{B}$ never exceeds the classical boundary $\mathcal{B}=2$. Thus, a quantum mediator prepared in a \textit{classical} initial state \textit{cannot} induce a violation of the temporal Bell inequality for the quantum probe $Q$, meaning that it cannot induce a non-classical dynamical evolution on $Q$ capable of entangling it at times $t_0$ and $t>t_0$.

This gives an important novel feature to the temporal witness of non-classicality. Observing a quantum coherent evolution of the quantum probe $Q$ mediated by the unknown system $M$ under global energy-conserving dynamics is a sufficient condition to show that $M$ must have been prepared in a \textit{non-classical} state, i.e., an eigenstate of a variable that does not commute with the classical one. Since one can in principle prepare $M$ in an eigenstate of this variable, we can qualify it as an \textit{observable}.

\textbf{Non-Classical initial state of $\mathbf{M}$}.-- We shall now consider the mediator $M$ to be initialised in a \textit{non-classical} state. While we present the most general result for $\mathcal{B}$ in this case in Appendix A, here we consider for simplicity the initial state to be an eigenstate of the observable {$q_y^M$: $\rho_M^{non-class}=\frac{1}{2}\left(\mathrm{I}+q_y^{M}\right)$. We set, without loss of generality and to inform the experiment, $a=0$, $b=0$, $c=1$, as we have shown in Eqs.~\ref{eq:Bclassicalmediator} and ~\ref{eq:Bellclassicalstate} that the classical parameters cannot lead to a violation of the temporal Bell inequality on their own.}

\begin{figure}
    \centering
    \includegraphics[width=\linewidth]{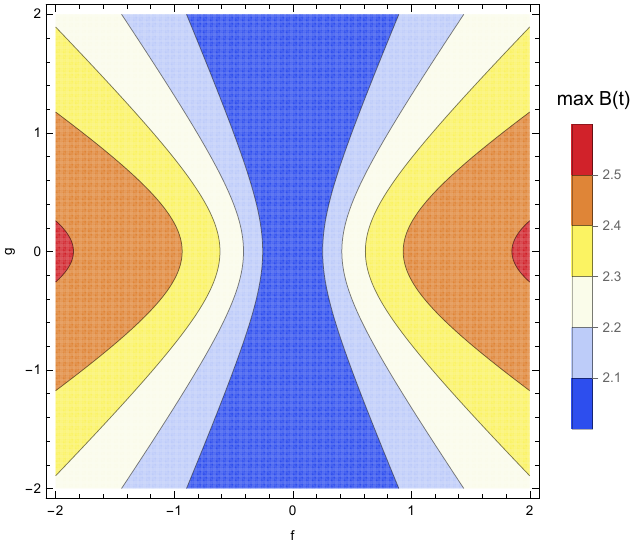}
    \caption{Contour plot of $\max_t{\mathcal{B}(t)}$ as a function of the non-classical parameters $f,\ g$. {The classical parameters are set to be $a=0,\ b=0,\ c=1$. $M$ is initialised in $\rho_M=\frac{1}{2}(\mathrm{I}+q_y^M)$.} The plot shows the existence of pairs $(f,g)$ determining a violation of the temporal Bell inequality with a non-classical $M$ initialised in a non-classical state $\rho_M$.}
    \label{fig:maxBnonClass}
\end{figure}

With these assumptions, we plot the maximum of the Bell quantity $\mathcal{B}$ for different values of the non-classical parameters $f,g\in\left[-2,2\right]$, as shown in Fig.~\ref{fig:maxBnonClass}.
The plot shows a clear violation of the classical limit $\mathcal{B}=2$ for almost all pairs $(f,g)$, meaning that a non-classical mediator prepared in a non-classical state \textit{can} induce temporal quantum correlations on the quantum probe $Q$, generating entanglement in time between $Q$ at time $t_0$ and at time $t>t_0$.

This allows us to restate the temporal witness of non-classicality as:
\begin{teorema}
    \textbf{Temporal entanglement-based witness of non-classicality} \\
    If $M$ can entangle \textit{in time} $Q$ at time $t_0$ and at time $t>t_0$, by mediating a global energy-conserving dynamics, then $M$ is non-classical.
    \label{theorem:temporalentanglement}
\end{teorema} 

\section{Experimental emulation}
We performed an emulation of the quantum network shown in Fig.~\ref{fig:CHSHscheme} using NMR \cite{Cory1997,Jones2024}. Because of the lack of true measurments it is necessary to store and combine partial results using an ancilla qubit \cite{Fitzsimons2015}, and so in this case a three spin NMR system is required. We used the three \nuc{19}{F} nuclei in a sample of iodotrifluoroethene \cite{Du2007,Katiyar2013} dissolved in acetone-d${}_6$. All quantum logic gates were implemented using shaped pulses designed using GRAPE \cite{Khaneja2005}, using phase-only control \cite{Violaris2021}. More details can be found in Appendix B.

\begin{figure}
\begin{center}
\includegraphics[]{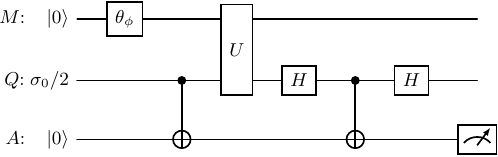}
\end{center}
\caption{A typical circuit to emulate temporal entanglement in a three-qubit NMR system; this example shows the case of a $\sigma_z$ measurement followed by a $\sigma_x$ measurement. For more details see the text.} \label{fig:NMRcircuit}
\end{figure}

A typical circuit for the NMR implementation is shown in Fig.~\ref{fig:NMRcircuit}. The ancilla qubit $A$ begins in state \ket{0}, while the mediator qubit $M$ starts in some pure state, prepared from \ket{0} using a $\theta_\phi$ single-qubit rotation. The quantum system qubit $Q$ begins in the maximally mixed state, $\sigma_0/2$. Quantum measurements of $\sigma_z$ are simulated using a simple \CNOT\ gate, while measurements of $\sigma_x$ are simulated using a \CNOT\ gate preceded and followed by Hadamard gates. Finally the measurement gate was simulated by applying a $90^\circ$ excitation pulse to spin $A$, and then observing the NMR spectrum.

\begin{figure}
\includegraphics{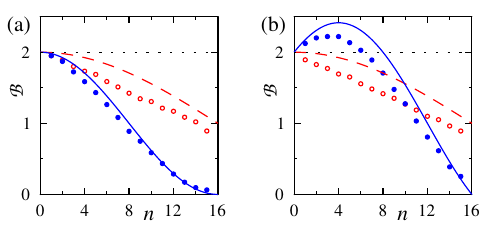}
\caption{Data from our emulation experiments. The lines indicate theoretical calculations while the points indicate experimental measurements, and red dashed lines/open points indicate a classical $M$ (mixing Hamiltonian with $f=0$), while blue solid lines/filled points indicate a non-classical $M$ (mixing Hamiltonian with $f=1$). Note that for the classical case the Bell inequality $\mathcal{B}$ never exceeds 2, indicated by the dotted line, while for the non-classical case deviations are seen when $M$ starts in a non-classical state, shown in (b), but not when it starts in a classical state, shown in (a). The observed damping in the experimental results, which causes these to deviate from theoretical calculations reflects a build up of errors arising from imperfect implementations of the logic gates.}\label{fig:data}
\end{figure}

Experiments were performed for two different initial states of $M$, for two different mixing Hamiltonians, and for a range of mixing times. The pseudo-pure state prepares $M$ in an effective state \ket{0}, which was used first directly (a classical pure state), and then after the application of a $90^\circ_x$ pulse (a non-classical pure state). The mixing Hamiltonians both had $a=b=g=0$ and $c=1$, but varied by choosing $f=0$ (simulating a classical $M$) or $f=1$ (a non-classical $M$). Experiments were performed for each of the four possible measurement combinations and for a range of mixing times, exhibiting both classical and non-classical behaviour. The individual measured integrals were normalised and then combined offline to produce the desired temporal Bell inequality, $\mathcal{B}$. The results are depicted in Fig.~\ref{fig:data}

\section{Discussion}
Theorem~\ref{theorem:temporalentanglement} represents the core result of this work, connecting witnesses of non-classicality in the temporal domain with the theory of entanglement in time.
It proves that $M$'s induced entanglement \textit{in time} is a sufficient condition for its non-classicality, namely the existence and dynamical involvement of (at least) two \textit{non-commuting observables}, provided that a global quantity on $Q\oplus M$ is conserved during the evolution.

This framework encompasses standard Leggett--Garg Inequalities (LGIs) tests, extending their applicability to multi-system configurations. Since the temporal Bell inequality in Eq.~\ref{eq:temporalBellineq} is mathematically equivalent to four-time LGIs \cite{emary_leggettgarg_2013}, our result also captures their violations. However, LGI-based tests rely on the assumption of \textit{non-invasive measurability}, introducing the ``clumsiness loophole", frequently invoked to challenge their validity as demonstrations of ``quantumness" \cite{wilde_addressing_2012}. Theorem~\ref{theorem:temporalentanglement} avoids this issue by using the temporal entanglement generated on a well-characterised auxiliary quantum system $Q$ to indirectly witness the non-classicality of $M$, of uncertain quantum nature.

Compared to Theorem~\ref{theorem:temporalold}, the use of temporal entanglement strengthens the witness: successful detection now requires $M$ to be initialised in a non-classical state -- i.e., an eigenstate of an observable non-commuting with the classical one. This excludes semi-classical models, such as classical fields inducing quantum coherence in $Q$, and implies that the observed temporal entanglement in $Q$ traces back to a prior non-classical preparation of $M$. Consequently, the non-commuting variables of $M$ function as true \textit{observables} of the system, as $M$ must have been prepared in their eigenstates for the witnessing to succeed -- an insight not accessible through the GWT.

Replacing quantum coherence in $Q$ (as in Theorem~\ref{theorem:temporalold}) with entanglement \textit{in time} as the witnessing task, our result offers a clear temporal counterpart to the GWT (Theorem~\ref{theorem:gwt}). This leads to several implications.

First, mediating two-fold quantum correlations, spatial or temporal, is sufficient to infer the non-classicality of the mediator, under appropriate assumptions. Thus, tools recently developed to quantify non-classicality in spatial scenarios \cite{raia_role_2024,krisnanda_revealing_2017,ganardi_quantitative_2024} can be adapted to the \textit{temporal} case by replacing spatial entanglement measures with suitable temporal ones. This sets desirable mathematical conditions (e.g., \textit{gd}-continuity \cite{ganardi_quantitative_2024}) for temporal entanglement measures.

Second, our framework simplifies experimental implementations in contexts like gravity and biology, where the GWT is challenged by spatial locality requirements \cite{di_pietra_bose-marletto-vedral_2024,krisnanda_probing_2018}. By witnessing non-classicality via temporal entanglement using a \textit{single} quantum probe, our approach bypasses the need for two non-directly-interacting probes, making Theorem~\ref{theorem:temporalentanglement} particularly practical.

Third, the symmetry between the witnessing tasks in the GWT and in our temporal witness extends to their \textit{assumptions}. In the GWT, \textit{the principle of locality} ensures that spatial entanglement between $Q$ and $Q'$ is due to the mediator only \cite{di_pietra_role_2024}. In our case, \textit{conservation of a global quantity} on $Q\oplus M$ plays an analogous role, preventing spontaneous evolution of $Q$ from determining temporal entanglement irrespective of $M$ \cite{brukner_quantum_2004, fritz_quantum_2010}.
This suggests a definition of locality in time: interactions are \textit{local in time} if they generate an evolution that conserves a global quantity at \textit{every} instant. This definition is \textit{model-independent}, as it relies on a general principle such as the conservation of the global observable, and aligns with earlier proposals \cite{brukner_quantum_2004,adlam_spooky_2018}. It also offers a clear criterion for \textit{temporal nonlocality}: an unsuccessful temporal entanglement-based witness of non-classicality occurring despite the presence of a conserved global quantity would constitute evidences for nonlocality in time \cite{letertre_temporal_2025}. 

Lastly, unlike the GWT (Theorem~\ref{theorem:gwt}) \cite{marletto_witnessing_2020}, Theorem~\ref{theorem:temporalentanglement} assumes the \textit{quantum formalism}, via its conservation-law formulation. However, conservation laws can also be framed in \textit{counterfactual} terms within Constructor Theory \cite{marletto_constructor_2016,marletto_information-theoretic_2022}, enabling a dynamics-independent version of our result, potentially applicable to post-quantum theories such as superinformation theories.

We shall leave these investigations to future works.
\\
\\
\textbf{Acknowledgments}. We thank Simone Rijavec and Nicetu Tibau Vidal for helpful discussions on a previous version of this manuscript. G.D.P thanks the Clarendon Fund and the Oxford Thatcher Graduate Scholarship for supporting this research. This publication was made possible through the support of the Gordon and Betty Moore Foundation, of the Eutopia Foundation, of B. Vass, and of the ID 61466 grant from the John Templeton Foundation, as part of the The Quantum Information Structure of Spacetime (QISS) Project (qiss.fr). The opinions expressed in this publication are those of the authors and do not necessarily reflect the views of the John Templeton Foundation. AJB has received funding from the European Research Council (ERC) under the European Union’s Horizon 2020 research and innovation programme (grant agreement No 101002859).
\bibliographystyle{apsrev4-2}
\bibliography{temporal}

\end{document}


\onecolumngrid

\renewcommand{\theequation}{S\arabic{equation}}
\setcounter{equation}{0}

\begin{center}
  \large \textbf{Supplementary Material for "Temporal Entanglement and Witnesses of Non-Classicality"}
\end{center}

\section{A.\ \ \ \ General Expression for $\mathcal{B}$}
We shall here include the analytical expressions for $\mathcal{B}$ in the general case where both the mediator $M$ and the quantum probe $Q$ are qubits. 
The protocol to compute $\mathcal{B}$, visually described in Fig.~\ref{fig:CHSHscheme}, requires choosing two observables $A_{1,2}$ to be measured at time $t_0$, let them evolve following the energy-conserving dynamics generated by the Hamiltonian in Eq.~\ref{eq:HamiltonianNonClassical}, and finally measuring the time-evolved variables $B_{1,2}=U^\dagger A_{1,2}U$. We recall here that all the measurements are performed on the quantum probe $Q$ only, as $M$ is the system under investigation and, thus, inaccessible.
Since the algebra of observables of $Q$ is generated by the vector of descriptors $\hat{\mathbf{q}}^Q=\left\{q_x^Q(t_0),q_y^Q(t_0),q_z^Q(t_0)\right\}=\left\{\sigma_x\otimes\mathrm{I},\sigma_y\otimes\mathrm{I},\sigma_z\otimes\mathrm{I}\right\}$, it is enough to provide the time evolved descriptors $q_z^Q(t)$ and $q_x^Q(t)$ to construct all the possible observables of the quantum probe $Q$ (we label $q_i^Q(t_0)\equiv q_i^Q,\ i=x,y,z$ to ease the notation):

\begin{align}
    q_z^Q(t)=&\frac{(a-b)^2+4(f^2+g^2)\cos^2{\sqrt{(a-b)^2+4(f^2+g^2)}}}{(a-b)^2+4(f^2+g^2)}q_z^Q +
    \frac{4(f^2+g^2)\sin^2{\left(\sqrt{(a-b)^2+4(f^2+g^2)}t\right)}}{(a-b)^2+4(f^2+g^2)}q_z^M & \nonumber \\ &
    -\frac{\sin{\left(2\sqrt{(a-b)^2+4(f^2+g^2)}t\right)}}{\sqrt{(a-b)^2+4(f^2+g^2)}}\left[f\left(q_x^Qq_y^M-q_y^Qq_x^M\right)-g\left(q_x^Qq_x^M+q_y^Qq_y^M\right)\right] & \nonumber \\ &
    +\frac{(a-b)\left[1-\cos{\left(2\sqrt{(a-b)^2+4(f^2+g^2)}t\right)}\right]}{(a-b)^2+4(f^2+g^2)}\left[g\left(q_x^Qq_y^M-q_y^Qq_x^M\right)+f\left(q_x^Qq_x^M+q_y^Qq_y^M\right)\right]
    \label{eqapp:qzt}
\end{align}
and
\begin{align}
    q_x^Q(t)=&\frac{\sin{\left(\sqrt{(a-b)^2+4(f^2+g^2)}t\right)}}{\sqrt{(a-b)^2+4(f^2+g^2)}}\left(\cos{\left[\left(a+b+2c\right)t\right]}+\cos{\left[\left(a+b-2c\right)t\right]}\right)\left(fq_z^Qq_y^M-gq_z^Qq_x^M\right) & \nonumber \\ &
    -\sin{\left[\left(a+b+2c\right)t\right]}\left(q_y^M+q_z^Qq_y^M\right) -\sin{\left[\left(a+b-2c\right)t\right]}\left(q_y^M-q_z^Qq_y^M\right) & \nonumber \\ &
    +\left[\cos{\left(\sqrt{(a-b)^2+4(f^2+g^2)}t\right)}\left(q_x^Q-q_x^Qq_z^M\right)+\frac{(a-b)\sin{\left(\sqrt{(a-b)^2+4(f^2+g^2)}t\right)}}{\sqrt{(a-b)^2+4(f^2+g^2)}}\left(q_y^Qq_z^M-q_y^Q\right)\right]\cdot & \nonumber \\ &
    \cdot\frac{1}{4}\left[\cos{\left((a+b-2c)t\right)}\left(q_x^Q-q_x^Qq_z^M\right)-\sin{\left((a+b-2c)t\right)\left(q_y^Q-q_y^Qq_z^M\right)}\right]  & \nonumber \\ &
    +\left[\cos{\left(\sqrt{(a-b)^2+4(f^2+g^2)}t\right)}\left(q_x^Q+q_x^Qq_z^M\right)-\frac{(a-b)\sin{\left(\sqrt{(a-b)^2+4(f^2+g^2)}t\right)}}{\sqrt{(a-b)^2+4(f^2+g^2)}}\left(q_y^Qq_z^M+q_y^Q\right)\right]\cdot & \nonumber \\ &
    \cdot\frac{1}{4}\left[\cos{\left((a+b+2c)t\right)}\left(q_x^Q+q_x^Qq_z^M\right)-\sin{\left((a+b+2c)t\right)\left(q_y^Q+q_y^Qq_z^M\right)}\right].
    \label{eqapp:qxt}
\end{align}

We shall consider $A_1=q_z^Q$, $A_2=q_x^Q$, so that $B_1=U^\dagger q_z^Q U$ and $B_2=U^\dagger q_x^Q U$ are given by Eqs.~\ref{eqapp:qzt} and ~\ref{eqapp:qxt} respectively.

To compute the temporal quantum correlations $E(A_i,B_j),\ i,j=1,2$ as in Eq.~\ref{eq:corrfunctionhybrid}, we need to specify the initial state of the mediator $M$. We shall choose here the most general: $\rho_M=\frac{1}{2}\left(\mathrm{I}+\alpha q_x^M+\beta q_x^M+\gamma q_x^M\right)$ with $\alpha,\beta,\gamma \in \mathbb{R}$ such that $|\alpha|^2+|\beta|^2+|\gamma|^2=1$.

The final, most general Bell quantity $\mathcal{B}$ thus reads:

    \begin{align}
    &\mathcal{B}=\left| -\frac{2 \cos (2 c t) \sin \left(\sqrt{(a-b)^2+4 \left(f^2+g^2\right)}t\right)\left[\sin\left((a+b)t\right) (\alpha  f+\beta  g)+\cos\left((a+b)t\right) (\beta  f-\alpha  g)\right]}{\sqrt{(a-b)^2+4 \left(f^2+g^2\right)}} \right. & \nonumber \\ & \left. +\frac{1}{2} \left[\frac{(a-b)\left[(\gamma -1) \sin\left((a+b-2 c)t\right)-(\gamma +1) \sin\left((a+b+2 c)t\right)\right] \sin \left(\sqrt{(a-b)^2+4 \left(f^2+g^2\right)}t\right)}{\sqrt{(a-b)^2+4 \left(f^2+g^2\right)}} \right. \right. & \nonumber \\ & \left. \left. +(\gamma +1) \cos\left((a+b+2 c)t\right) \cos \left(\sqrt{(a-b)^2+4 \left(f^2+g^2\right)}t\right) -(\gamma -1) \cos\left((a+b-2 c)t\right) \cos \left(\sqrt{(a-b)^2+4 \left(f^2+g^2\right)}t\right)\right] \right. & \nonumber \\ & \left. +2 \sin \left(t \sqrt{(a-b)^2+4 \left(f^2+g^2\right)}\right) \left(\frac{(a-b) (\alpha  f+\beta  g) \sin \left(t \sqrt{(a-b)^2+4 \left(f^2+g^2\right)}\right)}{(a-b)^2+4 \left(f^2+g^2\right)} \right. \right. \nonumber & \\ & \left. \left. +\frac{(\alpha  g-\beta  f) \cos \left(t \sqrt{(a-b)^2+4 \left(f^2+g^2\right)}\right)}{\sqrt{(a-b)^2+4 \left(f^2+g^2\right)}}\right)+\frac{(a-b)^2+4 \left(f^2+g^2\right) \cos ^2\left( \sqrt{(a-b)^2+4 \left(f^2+g^2\right)}t\right)}{(a-b)^2+4 \left(f^2+g^2\right)}\right|.
    \label{eq:Bmostgeneralcase}
\end{align}

By manipulating this expression, one can explore the possible regimes for the hybrid system:
\begin{enumerate}
    \item \textbf{Classical Mediator}: Set $f=g=0$. A classical system is a system characterised by a single variable, as all variables commute. The classical variable, identified by the conservation law, is $q_z^M$. This implies that $\alpha=\beta=0$ and leaves $\gamma$ undetermined, with the condition $|\gamma|^2\leq1$. This is the case we analysed in Eq.\ref{eq:Bclassicalmediator}: irrespectively of $\gamma$, a classical mediator can never mediate entanglement in time on the quantum probe $Q$;
    \item \textbf{Classical $\mathbf{M}$'s initial state}: Set $\alpha=\beta=0$. A classical initial state for $M$ does not imply that $M$ itself must be a classical system; thus, this sets no conditions on $f$ and $g$. Modifying $\gamma$, one can explore different classical states for $M$, defined as eigenstates of the classical variable ($q_z^M$) or mixed states: the former case corresponds to $\gamma=\pm1$ (Eq.~\ref{eq:Bellclassicalstate}), the latter to $\gamma=0$;
    \item \textbf{Non-classical $\mathbf{M}$'s initial state}: This general case doesn't restrict the choice of any parameter. However, cases 1. and 2. show that the parameters $a,b,c$, responsible for the classical contribution to the Hamiltonian in Eq.~\ref{eq:HamiltonianNonClassical}, cannot cause a violation of the temporal Bell inequality alone: Eq.~\ref{eq:Bclassicalmediator} and ~\ref{eq:Bellclassicalstate} show that $\mathcal{B}\leq 2$ when $f=g=0$, so in the absence of non-classical contributions to the Hamiltonian. 
    The generality of Eq.~\ref{eq:Bmostgeneralcase} can guide experimentalists in choosing the most appropriate observables to measure on the quantum probe $Q$ to determine the $M$-induced temporal entanglement ($\alpha,\ \beta,\ \gamma$), or to tune the interactions between $Q$ and $M$ in order to maximise the violation of the temporal Bell inequality ($f,\ g$). If the above parameters are set, Eq.~\ref{eq:Bmostgeneralcase} can tell what is the right amount of interaction time $t$ between $Q$ and $M$ to observe the appropriate violation of the inequality.
    Specifically, to retrieve the case discussed in the main text one should set $a=0,\ b=0,\ c=1$ and $\alpha=\gamma=0,\ \beta=1$.
\end{enumerate}

In conclusion, Eq.~\ref{eq:Bmostgeneralcase} contains all the physical scenarios one can obtain when considering a hybrid system $Q\oplus M$ obeying the assumptions of Theorem~\ref{theorem:temporalold}, namely (i) conservation of a global observable and (ii) the formalism of quantum theory. It allows for a complete study of the phenomenon of $M$-induced temporal entanglement and gives robustness to the connection between the temporal witness of non-classicality and the theory of temporal entanglement, which is summarised in Theorem~\ref{theorem:temporalentanglement}, the main result of this work.

\section{B.\ \ \ \ NMR details}
All NMR experiments were perfomed on an Agilent Inova 600 NMR spectrometer, with a \nuc{19}{F} resonance frequency of approximately 564\,MHz and at a temperature of 298\,K. The offset frequencies in iodotrifluoroethene dissolved in acetone-d${}_6$ were $\nu_M=14186.66$\,Hz, $\nu_Q=4.13$\,Hz, and $\nu_A=-20711.56$\,Hz, with spin--spin couplings of $J_{MQ}=69.77$\,Hz, $J_{MA}=47.85$\,Hz, and $J_{QA}=-128.24$\,Hz. Approximate relaxation times were $T_1\approx2.9$\,s and $T_2\approx0.11$\,s for the three spins.

All quantum logic gates were implemented using shaped pulses designed using GRAPE \cite{Khaneja2005}, using phase-only control \cite{Violaris2021}, with a fixed nominal amplitude of 25\,kHz and phase steps of 2\,$\mu$s, and with a tolerance of RF inhomogeneity achieved by averaging over RF field strengths between 95\% and 103\% of the nominal value. To avoid the danger of over-compilation \cite{Smolin2013} separate pulses were designed for each gate in Fig.~\ref{fig:NMRcircuit}, except that the $\sigma_x$ measurement was implemented as a single pulse implementing the sequence of three gates shown. For the $U$ gate, indicating the interaction between M and Q, a pulse was designed for each value of $f$ for a fixed evolution angle, and larger angles achieved by applying this evolution pulse repeatedly. All GRAPE pulses had a length between 1 and 7\,ms, with longer pulses for the controlled-gates which require evolution under the spin--spin couplings.

As in almost all NMR quantum information implementations it is necessary to begin by preparing a pseudo-pure state, also known as an effective pure state \cite{Cory1997,Gershenfeld1997}. This was achieved using a simplified form of the controlled-transfer gate approach \cite{Kawamura2010}. At thermal equilibrium the three spins all have the same polarisation, but this can be converted to the ratios 1:0:2 by applying a $60^\circ$ pulse to spin M and a $90^\circ$ pulse to spin Q, followed by a crush gradient. A pseudo-pure state of spins M and A can then be prepared using a single transfer gate \cite{Kawamura2010,Cory1998a}.

\bibliographystyle{apsrev4-2}
\bibliography{temporal}